\documentclass[fleqn,usenatbib,table,xcdraw]{mnras}

\usepackage[T1]{fontenc}

\DeclareRobustCommand{\VAN}[3]{#2}
\let\VANthebibliography\thebibliography
\def\thebibliography{\DeclareRobustCommand{\VAN}[3]{##3}\VANthebibliography}
\newcommand{\borg}{{\sc borg}}
\newcommand{\Msun}{\mathrm{M}_\odot}

\usepackage{graphicx}
\usepackage{orcidlink}
\usepackage{amsmath}
\usepackage{amssymb}
\usepackage{url}
\usepackage{newtxtext,newtxmath}

\title[The GLADE+ Galaxy Catalogue]{GLADE+: An Extended Galaxy Catalogue for Multimessenger Searches with Advanced Gravitational-wave Detectors}

\author[Dálya et al.]{
G. Dálya$^{1,2}$ \orcidlink{0000-0003-3258-5763}, R. Díaz$^{1}$, F. R. Bouchet$^{3}$ \orcidlink{0000-0002-8051-2924}, Z. Frei$^{1,4}$ \orcidlink{0000-0002-0181-8491}, J. Jasche$^{5}$ \orcidlink{0000-0002-4677-5843}, \newauthor G. Lavaux$^{3}$ \orcidlink{0000-0003-0143-8891}, R. Macas$^{6}$ \orcidlink{0000-0002-6096-8297}, S. Mukherjee$^{7}$ \orcidlink{0000-0002-3373-5236}, M. Pálfi$^{1}$ \orcidlink{0000-0001-5942-0470}, R. S. de Souza$^{8}$ \orcidlink{0000-0001-7207-4584}, \newauthor B. D. Wandelt$^{3,9,10}$ \orcidlink{0000-0002-5854-8269}, M. Bilicki$^{11}$ \orcidlink{0000-0002-3910-5809} and P. Raffai$^{1,4}$ \orcidlink{0000-0001-7576-0141}
\\
$^{1}$Institute of Physics, Eötvös Loránd University, 1117 Budapest, Hungary\\
$^{2}$Universiteit Gent, B-9000 Ghent, Belgium\\
$^{3}$ Institut d'Astrophysique de Paris, UMR 7095, CNRS, Sorbonne Universit\'e, 98bis Boulevard Arago, 75014 Paris, France\\
$^{4}$MTA-ELTE Astrophysics Research Group, 1117 Budapest, Hungary\\
$^{5}$The Oskar Klein Centre, Department of Physics, Stockholm University, AlbaNova University Centre,
SE 106 91 Stockholm, Sweden\\
$^{6}$University of Portsmouth, Institute of Cosmology and Gravitation, Portsmouth PO1 3FX, United Kingdom\\
$^{7}$ Perimeter Institute for Theoretical Physics, 31 Caroline Street N., Waterloo, Ontario, N2L 2Y5, Canada\\
$^{8}$Key Laboratory for Research in Galaxies and Cosmology, Shanghai Astronomical Observatory,\\ Chinese Academy of Sciences, 80 Nandan Rd., Shanghai 200030, China\\
$^{9}$ Sorbonne Universit\'e, Institut Lagrange de Paris,  98 bis Boulevard Arago, 75014 Paris, France\\
$^{10}$ Center for Computational Astrophysics, Flatiron Institute, 162 5th Avenue, 10010, New York, NY, USA\\
$^{11}$Center for Theoretical Physics, Polish Academy of Sciences, al. Lotników 32/46, 02-668 Warsaw, Poland\\
}

\date{Accepted XXX. Received YYY; in original form ZZZ}

\pubyear{2021}

\begin{document}
\label{firstpage}
\pagerange{\pageref{firstpage}--\pageref{lastpage}}
\maketitle

\begin{abstract}
We present GLADE+, an extended version of the GLADE galaxy catalogue introduced in our previous paper for multimessenger searches with advanced gravitational-wave detectors. GLADE+ combines data from six separate but not independent astronomical catalogues: the GWGC, 2MPZ, 2MASS XSC, HyperLEDA, and WISExSCOSPZ galaxy catalogues, and the SDSS-DR16Q quasar catalogue. To allow corrections of CMB-frame redshifts for peculiar motions, we calculated peculiar velocities along with their standard deviations of all galaxies having $B$-band magnitude data within redshift $z=0.05$ using the "Bayesian Origin Reconstruction from Galaxies" formalism. GLADE+ is complete up to luminosity distance $d_L=47^{+4}_{-2}$ Mpc in terms of the total expected $B$-band luminosity of galaxies, and contains all of the brightest galaxies giving 90\% of the total $B$-band and $K$-band luminosity up to $d_L\simeq 130$ Mpc. We include estimations of stellar masses and individual binary neutron star merger rates for galaxies with $W1$ magnitudes. These parameters can help in ranking galaxies in a given gravitational wave localization volume in terms of their likelihood of being hosts, thereby possibly reducing the number of pointings and total integration time needed to find the electromagnetic counterpart.
\end{abstract}
% Considering the brightest galaxies giving half of the total luminosity in the $W1$-band, GLADE+ has nearly 90 percent completeness even at $d_L\simeq 500$~Mpc.


\section{INTRODUCTION}

During their third observing run (O3), the network of advanced gravitational-wave (GW) detectors consisting of the two interferometers of Advanced LIGO \citep{Aasietal2015}, Advanced Virgo \citep{Acernese_2014}, and the recently joined \mbox{KAGRA} \citep{akutsu2020overview} issued tens of prompt public alerts of significant GW events to allow for follow-up observations with electromagnetic (EM) and neutrino observatories \citep{GraceDBwebsite,Abbott_2019}. Each alert includes the posterior probability distribution of the source sky position, for which the $90$ percent credible localisation area is typically a few hundred square degrees large \citep{2018LRR....21....3A}. For compact binary coalescence (CBC) events, luminosity distances can also be inferred, and thus full 3D sky localisation maps are released \citep{Abbott_2019}. Such events include suspected binary neutron star (BNS) mergers, which are the leading candidates for joint GW+EM observations \citep[see e.g.][]{Abbott_2017}.

An effective and typical way to find an EM counterpart for a GW event is to target a ranked set of potential host galaxies within the 3D localisation volume with follow-up telescope observations, and to look for variations between time-separated images of them \citep{2012A&A...539A.124L, 2013ApJ...767..124N, 2014ApJ...784....8H, 2014ApJ...795...43F, 2016ApJ...820..136G, 2016ApJ...829L..15S}. Besides telescope-dependent technical considerations and the probability distribution of the source location, one can also take into account astrophysically motivated factors in the ranking of host candidates. For example when dealing with a BNS merger event, one can sort host galaxy candidates by a model-based estimation of the BNS merger rate in them, which is expected to correlate with the star formation rate \citep{1991ApJ...380L..17P} (and thus with the blue luminosity, see e.g. \citealt{2012PhRvD..85j3004B}) and/or with the stellar mass \citep{2019MNRAS.487.1675A, 2020MNRAS.491.3419A} of the galaxies. In the favourable case of identifying the host galaxy of a GW event through observing its EM counterpart \citep[see e.g.][]{2017ApJ...848L..12A, 2017Sci...358.1556C}, one can refine the parameter estimations for the GW source with priors derived from the EM counterpart and properties of the host \citep[see e.g.][]{2017ApJ...851L..36G, 2017ApJ...851L..45G, 2019PhRvX...9a1001A}, or use GW parameter estimations to draw conclusions on the EM source (e.g. \citealt{2018PhRvL.121p1101A, 2018ApJ...852L..25R, 2019EPJA...55...50R}), the host (e.g. \citealt{2020ApJ...905...21A}), or on cosmological parameters including the Hubble constant (see \citealt{2017Natur.551...85A, 2021ApJ...909..218A, 2021arXiv211103604T}). Host galaxy candidates associated to detected CBC events can also be utilized in dark siren measurements of the Hubble constant when no EM counterpart of one or more CBC GW events is found \citep{2019ApJ...871L..13F, 2019ApJ...876L...7S, 2020PhRvD.101l2001G}, in mapping out the expansion history \citep{Mukherjee:2018ebj, Mukherjee:2020hyn, Bera:2020jhx, Diaz:2021pem}, and in testing the general theory of relativity based on GW propagation \citep{Mukherjee:2019wcg, Mukherjee:2020mha}.  

In \citet{dalya2018glade}, we introduced the {\it Galaxy List for the Advanced Detector Era} (GLADE) value-added full-sky catalogue of galaxies \citep{2016yCat.7275....0D}, which since then has extensively been used by the LIGO-Virgo Collaboration \citep{2017Natur.551...85A, 2020ApJ...896L..44A, 2021ApJ...909..218A, 2021ApJ...915...86A} and others (see \citealt{dalya2018glade} for references, and e.g. \citealt{2019ApJ...881L..26L, 2020MNRAS.492.3904A, 2020MNRAS.497..726G, 2021ApJ...909..126K, 2021ApJ...912..128P, 2021PTEP.2021eA104S, 2019ApJ...871L..13F, 2021arXiv210112660F}) for the purposes mentioned above. Additionally, GLADE has been integrated into tools supporting optimal selections of target galaxies for follow-up observations \citep{2019arXiv190407335R, 2019MNRAS.489.5775C, 2020A&A...634A..32S, 2020ApJ...894..127W, 2020PASP..132j4501X}, and was used for identifying or extracting information on potential hosts of fast radio bursts (see e.g. \citealt{2019ApJ...885L..24C}), short gamma-ray bursts (see e.g. \citealt{2020MNRAS.492.5011D}), supernova shock breakouts \citep{2020ApJ...896...39A}, and other transient phenomena observed in X-ray \citep{2017MNRAS.471.4990C}, optical (see e.g. \citealt{2021ApJ...908..180A, 2021MNRAS.503.4838K}), and radio bands (see e.g. \citealt{2017A&A...597A..96R, 2021MNRAS.505.2966K}).

In this paper, we introduce an updated, extended, and improved version of GLADE, which we will refer to as GLADE+. GLADE+ contains $\sim22.5$ million galaxies and $\sim750$ thousand quasars (compared to $\sim3$ million galaxies and $\sim300$ thousand quasars in GLADE) as a result of cross-matching GLADE with the WISE$\times$SuperCOSMOS Photometric Redshift Catalogue (WISExSCOSPZ) and updating the quasar database of GLADE with the SDSS-DR16Q quasar catalogue.  
To allow corrections of CMB-frame redshifts for peculiar motions, we calculated peculiar velocities along with their standard deviatons of all galaxies having $B$-band magnitude data within redshift $z=0.05$ using the "Bayesian Origin Reconstruction from Galaxies" formalism \citep{Mukherjee:2019qmm}. We estimated stellar masses and binary neutron star merger rates for all GLADE+ galaxies with $W1$-band magnitude data, and included these in the catalogue. The aim of these improvements is to continue serving the purposes mentioned above and potentially serve new ones, especially in light of the ongoing preparations for the fourth observing run of the LIGO-Virgo-KAGRA network scheduled for 2022 \citep{2018LRR....21....3A}.

This paper is organized as follows. In Section \ref{sec:construction} we introduce the catalogues and methods we used to create GLADE+, including the peculiar velocity correction and the stellar mass and binary neutron star merger rate estimations. In Section \ref{sec:completeness} we quantify the completeness of GLADE+ based on the measured $B$- and $K_s$-band luminosities of galaxies. In Section \ref{sec:description} we describe the exact format of the catalogue, and in Section \ref{sec:conclusion} we draw our conclusions.

Throughout this paper we adopt a flat $\Lambda$CDM cosmology with the following parameters from the Planck 2018 results: $H_0=100h=67.66\ \mathrm{km}\ \mathrm{s}^{-1}\ \mathrm{Mpc}^{-1}$, $\Omega_\mathrm{M}=0.3111$, and $\Omega_{\Lambda}=0.6889$ \citep{2020A&A...641A...6P}.

\section{CONSTRUCTION OF THE GLADE+ CATALOGUE}
\label{sec:construction}

We have constructed GLADE+ from six separate but not independent\footnote{2MPZ was created by cross-matching 2MASS XSC, WISE and SuperCOSMOS and using an artificial neural network approach trained on several redshift surveys to derive the photometric redshifts, hence it is not independent from these catalogues; for detailes see \citealt{2MPZ}.} astronomical catalogues: the Gravitational Wave Galaxy Catalogue\footnote{\url{http://vizier.u-strasbg.fr/viz-bin/VizieR?-source=GWGC}} (GWGC, see \citealt{GWGC}), HyperLEDA\footnote{\url{http://leda.univ-lyon1.fr/}} \citep{HyperLEDA}, the 2 Micron All-Sky Survey Extended Source Catalog\footnote{\url{https://old.ipac.caltech.edu/2mass/}} (2MASS XSC, see \citealt{2000AJ....119.2498J} and \citealt{2MASS_XSC}), the 2MASS Photometric Redshift Catalog\footnote{\url{http://ssa.roe.ac.uk/TWOMPZ.html}} (2MPZ, see \citealt{2MPZ}), the WISExSCOS Photometric Redshift Catalogue\footnote{\url{http://ssa.roe.ac.uk/WISExSCOS.html}} (WISExSCOSPZ, see \citealt{WISExSCOS}), and the Sloan Digital Sky Survey quasar catalogue from the 16th data release\footnote{\url{https://www.sdss.org/dr16/algorithms/qso_catalog/}} (SDSS-DR16Q, see \citealt{SDSS_DR16_Q}). As we have used the first four of these catalogues in creating the GLADE galaxy catalogue, the relevant characteristics of these are summarized in \cite{dalya2018glade}. Note, that there are several other survey catalogues available offering deep digital observations, such as the DESI Legacy Survey \citep{2019AJ....157..168D}, Pan-STARRS (\citealt{2020ApJS..251....7F}, \citealt{2021MNRAS.500.1633B}) or SkyMapper \citep{2019PASA...36...33O}, which we plan to incorporate in future versions of the catalogue. In Section \ref{sec:crossmatch} we only describe the WISExSCOSPZ and SDSS-DR16Q catalogues, and discuss the cross-matching between GLADE and WISExSCOSPZ, as well as the results we obtained. In Section \ref{sec:pecvel} we describe the method we used for estimating peculiar velocities, and in Section \ref{stellarmass} we introduce the methods used to estimate the stellar masses of and BNS merger rates in the individual galaxies.

\subsection{Cross-matching}
\label{sec:crossmatch}

The WISExSCOSPZ catalogue was constructed by cross-matching the AllWISE full-sky release \citep{AllWISE} of the Wide-field Infrared Survey Explorer (WISE, see \citealt{WISE}), which is the most comprehensive survey of the mid-infrared sky, and the SuperCOSMOS Sky Survey (\citealt{SCOS}), the result of an automated scanning and digitizing of photographic plates from the United Kingdom Schmidt Telescope and the Palomar Observatory Sky Survey-II. WISExSCOSPZ contains $\sim20$ million galaxies with photometric redshifts calculated using an artificial neural network algorithm \citep{2004PASP..116..345C}. The redshifts have errors nearly independent of distance, with an overall accuracy of $\sigma_z/(1+z) \simeq 0.033$ \citep{WISExSCOS}. The catalogue contains magnitude information in the $B_J$ and $R_F$ bands from SuperCOSMOS \citep{SCOS2} and in the $W1$ and $W2$ WISE bands. In order to calculate the magnitudes of the galaxies in the Johnson-Cousins $B$-band, we used the color equations presented in \cite{SCOS2}.

We have created GLADE+ by cross-matching GLADE v2.4\footnote{The specifics of different GLADE versions are described on the GLADE website: \url{http://glade.elte.hu}} with the WISExSCOSPZ catalogue and then replacing the quasars with the newer set from SDSS-DR16Q and removing the globular clusters. We could not use the method described in \cite{dalya2018glade} for cross-matching GLADE with WISExSCOSPZ, as duplicate galaxies could not simply be found by their designations. Hence we used a resolution of 2 arcseconds, i.e. if a WISExSCOSPZ galaxy lied closer to a GLADE galaxy than this threshold, we treated them as being the same object and merged them. This distance threshold was motivated by the fact that 2 arcseconds is the maximal resolution in the WISExSCOSPZ catalogue and false positive associations start to dominate above this value. The order of magnitude of the threshold is also consistent with that of previous GLADE cross-matches, note however that in previous applications we could take other parameters of the galaxies into account as well, such as luminosity distances and $B$ magnitudes \citep{dalya2018glade}.

The GLADE catalogue incorporated the SDSS-DR12Q quasar catalogue, which, in GLADE+, we replaced entirely with the more recent and extended SDSS-DR16Q catalogue. This catalogue contains data for $\sim750,000$ quasars (including the $\sim300,000$ quasars published in SDSS-DR12Q), which makes it the largest selection of spectroscopically confirmed quasars to date.

\begin{figure*}
	\includegraphics[width=2\columnwidth]{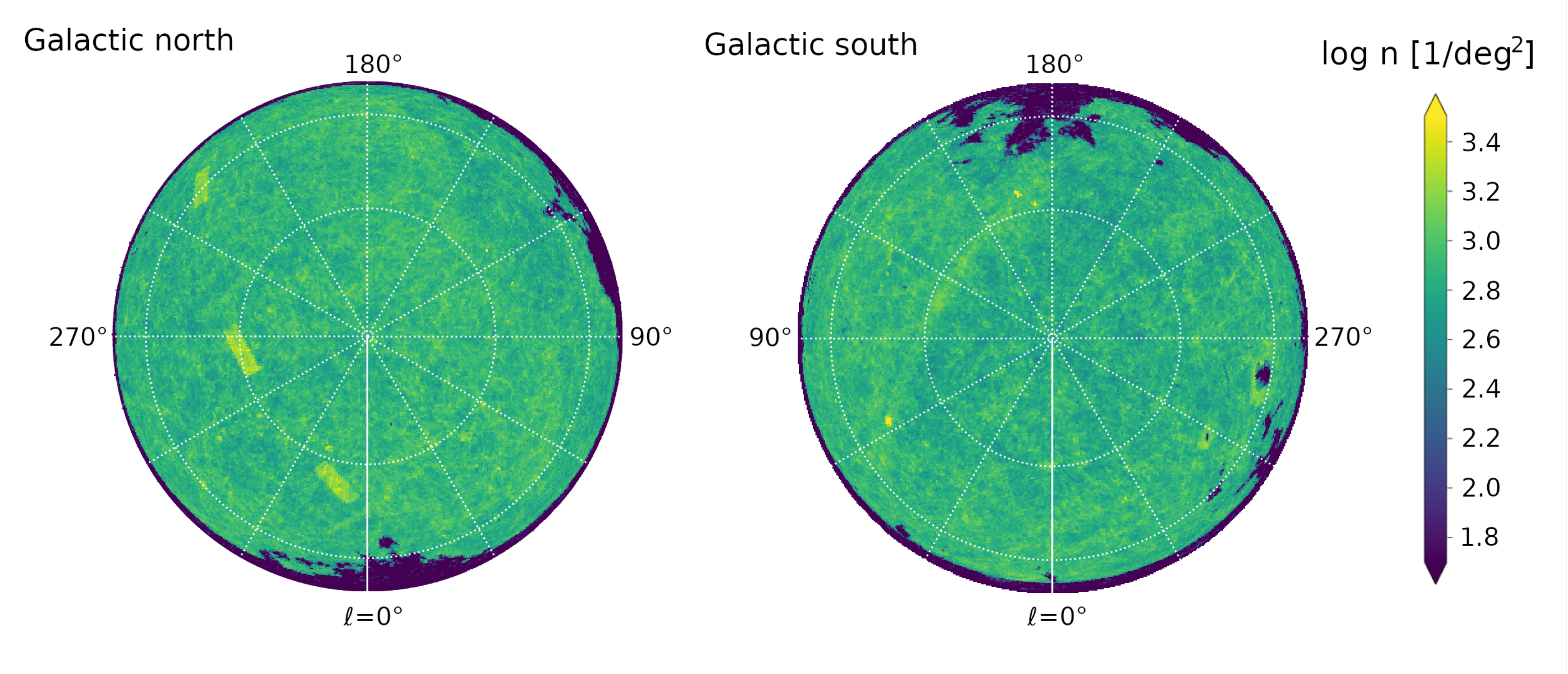}
    \caption{The base-10 logarithm of the number density ($n$) of objects in GLADE+ using azimuthal projection with galactic coordinates. The plane of the Milky Way obscures the visibility of background galaxies, hence the (blue) underdense regions at the edges of the plots. Overdense (yellow) patches and stripes originating from the HyperLEDA catalogue show up as a result of deeper, more sensitive surveys (such as SDSS and GAMA), that have been made towards the corresponding sky locations.}
    \label{fig:map}
\end{figure*}

Cross-matching and updating the catalogues resulted in the GLADE+ catalogue containing 23,181,758 objects from which 22,431,348 are galaxies and 750,410 are quasars. The sky distribution of GLADE+ objects are shown in Figure \ref{fig:map} as a density plot. The plane of the Milky Way is clearly noticeable in the figure, as the gas and dust reduces the visibility towards those directions and different sky surveys used various cuts in galactic latitude. Other anisotropies are arising from the different sensitivities and footprints of the various sky surveys.

Due to incorporating the WISExSCOSPZ catalogue, 21,165,400 galaxies in GLADE+ had $W1$ magnitudes available from that sample, which we used to estimate their stellar masses and the binary neutron star merger rates in them, see Section \ref{stellarmass}. In order to supplement as many of the remaining ones with $W1$ as possible, we first cross-matched them with AllWISE, using 3 arcsecond matching radius and keeping only the closest matches. We found a match for over 1.76 million galaxies, which left us with $\sim$260,000 without a WISE counterpart. For part of the latter (32,153 galaxies) we provide $W1$ apparent magnitude estimates based on $K_s$-band measurements available in 2MASS. To obtain these estimates, we first fitted the mean $W1-K_s$ colour as a function of redshift for sources with available spectroscopic redshifts, $K_s$ and $W1$ magnitudes in the 2MPZ catalogue. This effective $W1-K_s$ colour relation is then added to the $K_s$ band measurements for those sources in GLADE+ which have $K_s$ magnitudes from 2MASS but do not have the $W1$ ones, to obtain the $W1$ magnitude estimates.

\subsection{Peculiar velocity correction}
\label{sec:pecvel}
The correction of peculiar velocities for low redshift galaxies is essential to correctly calculate the true redshift.  
The estimation of peculiar velocities for galaxies in GLADE+ is made using the method proposed by \citet[][]{Mukherjee:2019qmm}, which relies on a Bayesian formalism called "Bayesian Origin Reconstruction from Galaxies" (\borg{}). The \borg{} forward modelling method infers a probabilistic and physically plausible model of the three-dimensional cosmic matter distribution from observed galaxies in cosmological surveys to derive the linear and partially the non-linear component of the velocity field \citep[see e.g.][]{jasche_bayesian_2013,jasche2015,lavaux_unmasking_2016,jasche2019}. This method solves a large-scale Bayesian inverse problem by fitting a dynamical structure formation model to data, and estimates the initial conditions of the early Universe from which presently-observed structures can be explained. 
The \borg{} algorithm marginalizes over unknown galaxy bias and accounts for selection and evolutionary effects while providing the velocity field as part of the dynamical model. The reliability of this method is verified with an $N$-body simulation to check the consistency of the velocity fields provided by the posterior distributions sampled by \borg{} and the one provided by the original $N$-body simulation \citep{Mukherjee:2019qmm}. This method gives a set of points in the parameter space (a spatial grid of $256^3$ values with a spatial resolution of $2.64$ Mpc $h^{-1}$ for the initial conditions plus the bias parameters) that provides a numerical approximation of the posterior distribution of these parameters given the observed large scale structure observation. For each sample of the posterior, initial and final positions of the dark matter particles are provided, from which the velocity field can be estimated using the Simplex-in-Cell estimator \citep[SIC,][]{Hahn2014_DMSHEET, Leclercq2017a}. More details on this method and its validation can be found in \cite{Mukherjee:2019qmm}.

Along with the velocity estimation from \borg{}, we also include the contribution from the non-linear virial component of the velocity field. The radial components of the virial velocities are modelled as Gaussian random variates with variance  \citep[][]{2001MNRAS.322..901S,Mukherjee:2019qmm}

 \begin{equation}\label{vvir}
\sigma_\text{vir}= 476\, g_v\, (\Delta_\text{nl}(z)E(z)^2)^{1/6} \ \left(\frac{M_{\mathrm{h}}}{10^{15}\, \Msun h^{-1}}\right)^{1/3},
\end{equation}
where $g_v=0.9$, $\Delta_\text{nl}(z)= 18\pi^2 +60x -32 x^2$, and $x= \Omega_\text{m} (1+z)^3/E^2(z)-1$; $E(z)$ is the cosmological expansion function. In order to use this relation, we need to estimate the halo mass $M_{\mathrm{h}}$. We have used a mass-luminosity relation \citep[][]{2004MNRAS.353..189V} 
\begin{equation}\label{lmr}
L_B = \frac{A(M_{\mathrm{h}}/M_r)^b}{[c+(M_{\mathrm{h}}/M_r)^{dk}]^{1/k}},
\end{equation}
where $A= 5.7\times 10^9$, $M_r=10^{11}\, \Msun$ is the parameter capturing the knee in the mass-luminosity relation, $b=4$, $c=0.57$, $d=3.72$, and $k=0.23$, to estimate the mass from the $B$-band luminosity $L_B$ of the galaxy. The total variance in the velocity field is then calculated as $\sigma^2_{\text{tot}}= \sigma^2_{\text{\borg{}}} + \sigma^2_{\text{vir}}$. 

Using this method we have estimated the mean value\footnote{The direction of the velocity field is chosen such that the positive value of the velocity field indicates that the object is moving away from us.} of the velocity field to all GLADE+ galaxies which cross-match with the 2M++ compilation \citep{Lavaux2011_TMPP,Mukherjee:2019qmm} for galaxies up to redshift $z=0.05$ for which the $B$-band luminosity is available, along with the standard deviation $\sigma_\mathrm{tot}$ of the peculiar velocity which includes both linear and non-linear components of the velocity field. The conversion from the heliocentric redshift to the CMB-frame redshift are performed using the observation of the CMB temperature anisotropy by FIRAS \citep{1996ApJ...473..576F}. The uncertainty in the velocity error is translated into an uncertainty in the redshift and is provided in the catalog.  

%Check https://arxiv.org/pdf/1804.05709.pdf page number 4, right side.

\subsection{Stellar mass and binary neutron star merger rate estimations}\label{stellarmass}
According to \citet{2019MNRAS.487.1675A}, stellar masses (i.e.~the total mass of the active and remnant stars) of galaxies strongly correlate with the merger rates of the colliding binaries. These parameters can help in ranking the galaxies in a GW localization volume for EM follow-up observations, thereby possibly reducing the number of pointings and the total integration time needed to find the EM counterpart. The stellar mass is also one of the key parameters of the formation and evolution of galaxies 
\citep[e.g.][]{Loon, Ahad, Engler}. Hence we aimed to estimate stellar masses and CBC merger rates for as many galaxies in GLADE+ as possible.

To estimate the stellar mass of a galaxy, the so-called (stellar) mass-to-light ratio ($M_\ast/L$) is required that can be obtained from stellar population synthesis models.
Then the estimation can be performed with spectral energy distribution fitting, or based on one or more magnitude bands or colours (see \citealt{Courteau} for a review).
Since only a few magnitude bands ($B$ and some infrared) are contained in GLADE+, and no spectral energy distributions are available from the source catalogues, it is straightforward to use one of the magnitude-based stellar mass estimation methods. Note, that spectral energy distributions could be obtained from external catalogues, with which more precise stellar mass estimations could be given; this is something we consider to do in later versions of our catalogue.
The mid-infrared $M_\ast/L$ is relatively insensitive to the different stellar populations, particularly in the absence of ongoing star formation, and in addition it is not very sensitive to dust attenuation \citep[see e.g.][]{Wen, Rock}. 
Therefore we estimated the stellar masses based on the mid-infrared WISE magnitudes.

According to \citet{Kettlety} the mass-to-light ratio in the $W1$-band is 
\begin{equation}\label{eq:mass-to-light}
M_\ast/L_{W1} = 0.65 \pm 0.07
\end{equation}
for passive galaxies, which, according to the authors' claim, can provide at least as accurate mass estimates for galaxies with redshifts $z \leq 0.15$ as other more complex methods.
For the galaxies with active star formation, they give the next equation:
\begin{equation}\label{eq:active}
    \log_{10} (M_\ast/L_{W1}) = -0.4 \pm 0.2.
\end{equation}
Here the stellar mass is in the usual solar mass ($M_\odot$) unit, and the WISE magnitudes are in the Vega system. We separated the passive and active galaxies: $W2-W3\leq 1.5$ indicates active star formation \citep{Cluver, Jarrett}. We identified 799,703 passive and 18,351,034 active galaxies. There are 2,628,585 galaxies having no galaxy type, we assumed that they are active. We calculated the stellar masses using Eq. \ref{eq:mass-to-light} \ref{eq:active} for galaxies having WISE $W1$ magnitude according to their galaxy type.  The $W1$ luminosities can be calculated as
\begin{equation}
L_{W1} (L_{\odot}) = 10^{-0.4(M -M_{\odot,W1} )},
\end{equation} where $M_{\odot,W1}=3.24^\text{m}$ is the $W1$ magnitude of the Sun and 
\begin{equation}
M = m+5-5 \log_{10} d_L - K
\end{equation}
is the absolute magnitude of the galaxy in the $W1$ band, $m$ is the apparent $W1$ magnitude, $d_L$ is the luminosity distance and
\begin{equation}
K = - 7.1 \log_{10}( 1 + z )
\end{equation} is the K correction of \citet{Kettlety}, where $z$ is the heliocentric redshift. We do not apply correction for extinction because WISE magnitudes are already corrected fro that \citep{Bilicki_2016}. 
We only accepted stellar masses larger than $10^5~M_\odot$ as this value is the lower limit of the stellar masses of dwarf  \citep[e.g.][]{Garrison-Kimmel_2019}. (There were no galaxies having stellar mass smaller than $10^5~M_\odot$.)
Using this method, we could estimate the stellar mass of 21,779,322 ($\sim97\%$) of the galaxies in GLADE+.

We also provide the error of the stellar mass of each galaxy using the propagation of uncertainty.
The errors of luminosity distances were calculated from the errors of redshifts in the cases where it is not known.
Where the error of the $W1$ magnitude was not known for an individual galaxy%\footnote{We took the $W1$ magnitude errors from \url{http://ssa.roe.ac.uk/www/ssa_browser.html}.}
, we used the mean $W1$ error from WISExSCOSPZ instead.
As a result of our calculations, we have found that the mean (median) relative stellar mass error is 40 (33) percent for passive galaxies and 67 (58) percent for active galaxies.
The codes of the stellar mass estimation are publicly available on the GLADE website. 

The BNS merger rates ($N_\text{BNS}$) can be calculated from the stellar mass values according to \citet{2019MNRAS.487.1675A}:
\begin{equation}\label{eq:Artale}
     \log_{10}(N_\text{BNS}/\text{Gyr}) = ( 1.15 \pm 0.08 ) \log_{10}(M_\ast/M_\odot) - ( 7.22 \pm 0.22 ),
\end{equation}
which is valid below $z \leq 0.1$ and for galaxies with stellar masses $M_\ast > 10^7 M_\odot$, so we calculated the BNS merger rates for galaxies satisfying these conditions. Applying these criteria resulted in 3,156,544 galaxies with binary neutron star merger rates.
In addition, \citet{2019MNRAS.487.1675A} provides equations for the merger rates of binary black holes: 
\begin{equation}
    \log_{10}(N_{\mathrm{BBH}}/\text{Gyr}) = ( 0.80 \pm 0.07 ) \log_{10}(M_*/M_{\odot}) - ( 4.14 \pm 0.19 ),
    \label{eq:BBHrate}
\end{equation}
and of black hole - neutron star pairs as well:
\begin{equation}
    \log_{10}(N_{\mathrm{NSBH}}/\text{Gyr}) = ( 0.87 \pm 0.08 ) \log_{10}(M_*/M_{\odot}) - ( 4.99 \pm 0.22 ).
    \label{eq:NSBHrate}
\end{equation}
Note, that we only provide merger rates for BNSs in GLADE+, however using the stellar masses from the catalogue and Equations \ref{eq:BBHrate}-\ref{eq:NSBHrate}, the binary black hole and the black hole - neutron star merger rates can be calculated as well.

\section{CATALOGUE COMPLETENESS}
\label{sec:completeness}

Following the methods we introduced in \cite{dalya2018glade}, we quantify the completeness of GLADE+ using two different methods: (i) by comparing, within different $d_L$ limits, the integrated $B$ luminosity of GLADE+ galaxies to calculated reference values, and (ii) by comparing luminosity distributions of GLADE+ galaxies to the Schechter function within different luminosity distance shells. 

\begin{figure*}
	\includegraphics[width=1.6\columnwidth]{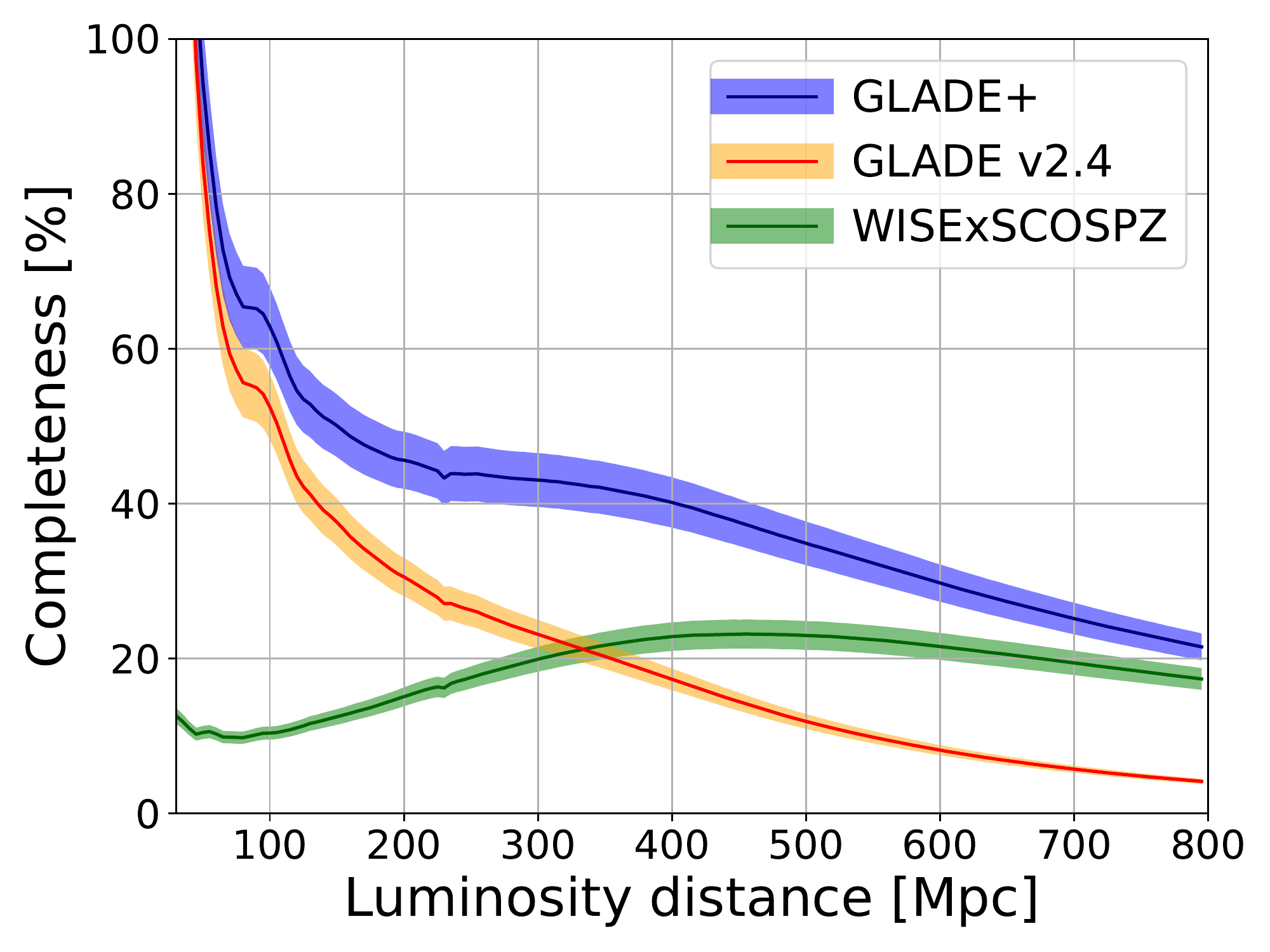}
    \caption{The completeness in terms of the normalized integrated $B$-band luminosity of galaxies in GLADE+ (blue) and its constituent catalogues (i.e.~GLADE v2.4 shown in orange and WISExSCOSPZ shown in green) within luminosity distances indicated on the $x$-axis. The normalization is carried out with the integrated $B$-band luminosity calculated from an average $B$-band luminosity density of a complete catalogue of homogeneously distributed galaxies (see \citealt{Kopparapu_2008} for details). The completeness value exceeds 100 percent within 47 Mpc due to a local overdensity of galaxies around the Milky Way. The completeness below $\sim 330$ Mpc is mostly due to GLADE galaxies, and WISExSCOSPZ galaxies contribute more for distances larger than $\sim 330$ Mpc.}
    \label{fig:completeness1}
\end{figure*}

The first method was originally used by \cite{GWGC} to calculate the completeness of the GWGC catalog. Here we compare the integrated $B$-band luminosity of GLADE+ galaxies within different luminosity distance limits to the total $B$-band luminosity we would expect from the same volume given a complete catalogue of homogeneously distributed galaxies with $B$-band luminosity density $(1.98 \pm 0.16) \times 10^{-2}\ L_{10}\ \mathrm{Mpc}^{-3}$, where $L_{10}=10^{10}\ L_{B,\odot}$ and $L_{B,{\odot}}$ is the solar luminosity in the $B$-band. Figure \ref{fig:completeness1} shows a comparison between the completeness values inferred using this method for GLADE+ and its constituent catalogues, GLADE v2.4 and WISExSCOSPZ. Completeness values over 100 percent are results of local overdensities of galaxies. The drop in completeness around $d_L\simeq 220$ Mpc corresponds to the distance limit for our peculiar velocity correction (see Section \ref{sec:pecvel}). As we can see from Figure \ref{fig:completeness1}, most of the completeness below $\sim 330$ Mpc comes from GLADE v2.4, and at larger distances contributions from WISExSCOSPZ galaxies start to dominate. Based on this completeness measure, GLADE+ is complete up to $d_L=47^{+4}_{-2}$ Mpc. GLADE+ has a completeness of $\sim$55 percent within the single-detector LIGO Livingston BNS range during O3 (130 Mpc) and $\sim$45 percent within the maximal planned single-detector BNS range in O4 (190 Mpc, see \citealt{2020LRR....23....3A}). Note, that the BNS range is the average distance from which a GW detector can detect a circular binary of two $1.4\ M_{\odot}$ neutron stars with a signal-to-noise ratio of 8, where the average is calculated over all possible sky positions and orbital inclinations, the maximum distance from which a BNS (i.e.~with optimal sky direction and inclination) can be detected is $\sim2.26$ times larger \citep{1993PhRvD..47.2198F}. Furthermore, the data-driven projections obtained by \cite{2022ApJ...924...54P} suggest that the median luminosity distance of BNS mergers observed during O4 will be 352.8 Mpc.

\begin{figure*}
	\includegraphics[width=2\columnwidth]{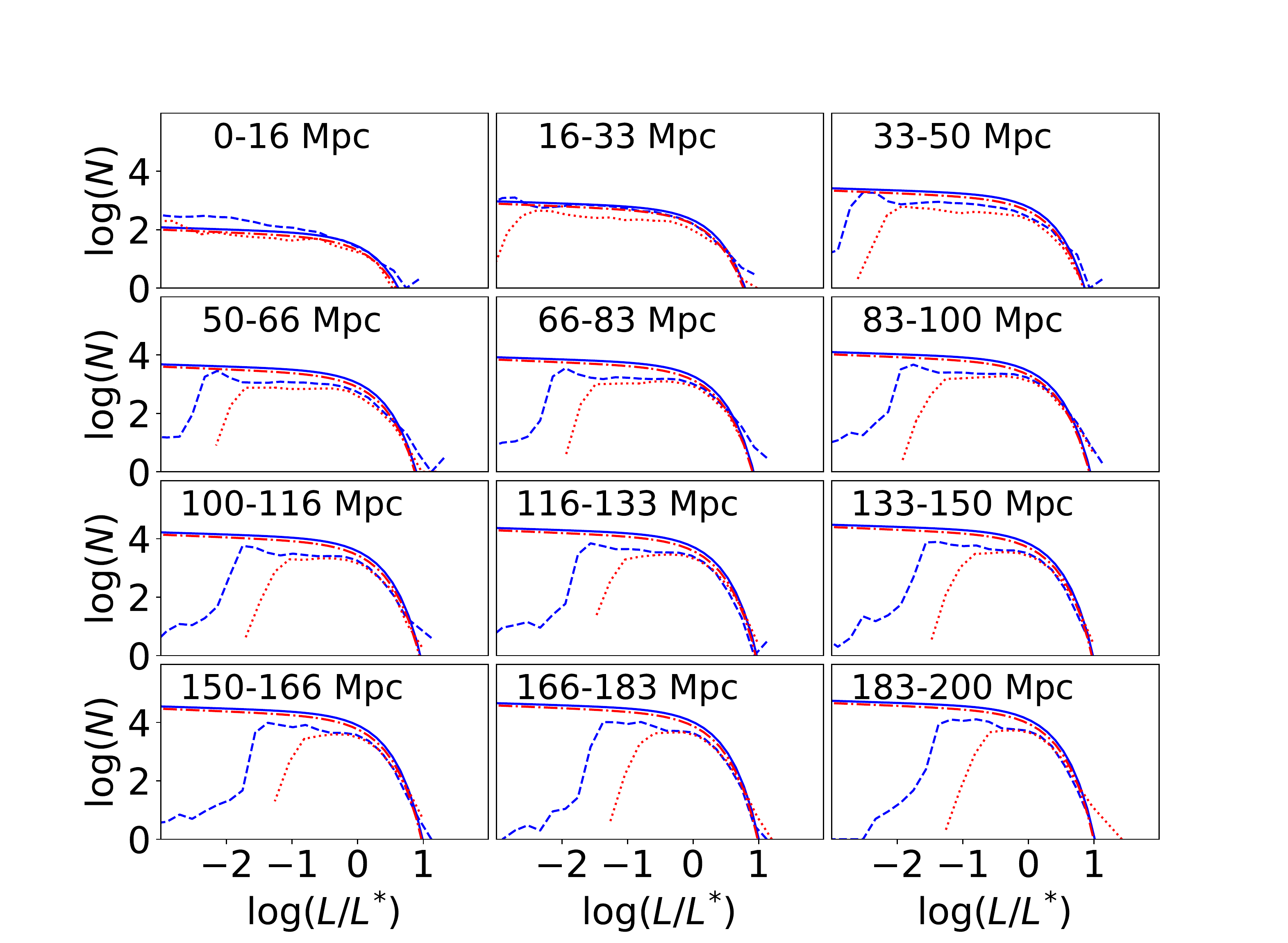}
    \caption{Luminosity histograms of GLADE+ galaxies within different luminosity distance shells in terms of their measured $B$-band and $K_s$-band luminosities (blue dashed and red dotted lines, respectively), compared to the same histograms we expect for complete catalogues based on $B$- and $K_s$-band Schechter functions (blue solid and red dash-dotted lines, respectively). $L^*$ is the characteristic luminosity of the Schechter function.}
    \label{fig:completeness2}
\end{figure*}

We have also used a second method to characterize the completeness of GLADE+. Here we divided the galaxies into 12 luminosity distance shells, each having a width of $\Delta d_L = 16.7\ \mathrm{Mpc}$. We have constructed histograms of $B$ and $K_s$ band luminosities of GLADE+ galaxies for each shell, which we show in Figure \ref{fig:completeness2} together with their corresponding Schechter functions. For the $B$-band Schechter function we have used the following values from \cite{2016ApJ...820..136G}: $\phi^*=1.6 \times 10^{-2}\ h^3\ \mathrm{Mpc}^{-3}$, $M_B^*=-20.47$ and $a=-1.07$. For the $K_s$-band Schechter function we have used the following parameters from \cite{2001ApJ...560..566K}: $\phi^*=1.16 \times 10^{-2}\ h^3\ \mathrm{Mpc}^{-3}$, $M_{K}^*=-23.39$ and $a=-1.09$. Figure \ref{fig:completeness2} shows that as distance increases, more and more faint galaxies are missing from GLADE+ in both bands. We can also see that the faint limit of our catalogue decreases more rapidly in the $K_s$-band. Note, that as Figure \ref{fig:completeness2} shows, even though the Schechter function fits quite well for the galaxy distributions in both the $B$ and the $K_s$ bands, it seems to systematically underestimate the number of the brightest galaxies. Hence, the completeness measure we can derive from it can only be an approximation of the catalogue's true completeness. 

%As by taking errors of these parameters also into account the two Schechter functions overlap, Figure \ref{fig:completeness2} shows only the one for the $B$-band. 

\begin{figure}
	\includegraphics[width=\columnwidth]{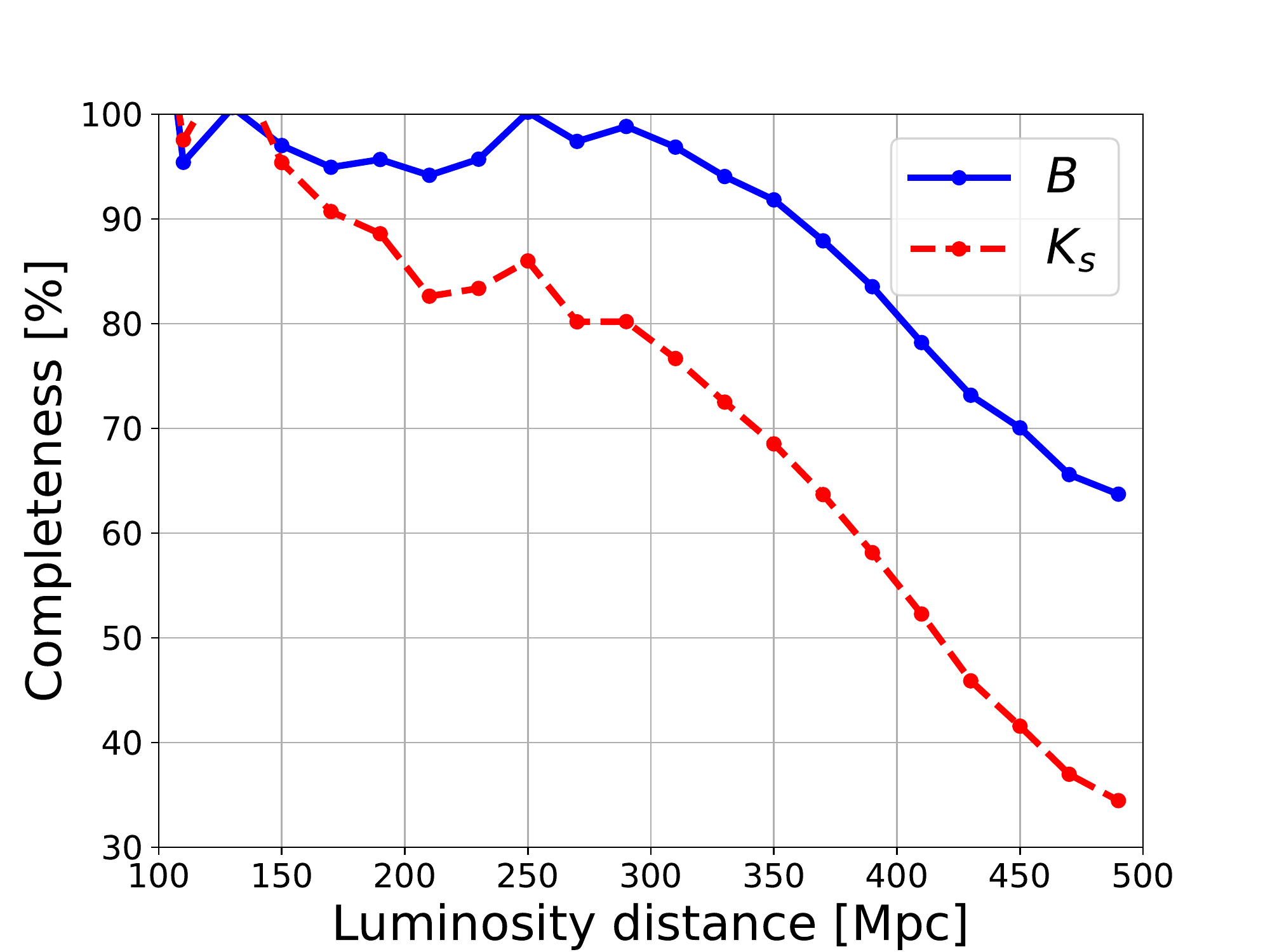}
   \caption{Completeness of GLADE+ in various distance shells having a 20 Mpc width relative to the $B$-band (blue solid line) and the $K_s$-band Schechter functions (red dashed line) for the the galaxies giving 90\% of the total luminosity in the given band. Note, that even if the completeness is 100 percent or above at a given distance, GLADE+ can still lack fainter galaxies.}
  \label{fig:completeness3}
\end{figure}

We have compared the integrated luminosity of a subset of galaxies giving 90\% of the total luminosity in each shell to the expected value corresponding to the Schechter function. In this analysis we have increased the luminosity distance limit to $d_L=500\ \mathrm{Mpc}$ and the shell width to $\Delta d_L=20\ \mathrm{Mpc}$. The completeness of GLADE+ in the $B$ and $K_s$-bands in the different bins is shown in Figure \ref{fig:completeness3}. The figure shows that the completeness for this subset of galaxies decreases more rapidly in the $K_s$-band over $d_L\simeq 100\ \mathrm{Mpc}$, and using this definition GLADE+ is complete up to $\sim 130\ \mathrm{Mpc}$ in the $B$ and $K_s$-bands. Note, that we have considered only 90\% of all the galaxies for this analysis, so even if the completeness is 100 percent or above at a given distance, GLADE+ can still lack a large number of the faintest galaxies there.

\section{DESCRIPTION OF THE GLADE+ CATALOGUE}
\label{sec:description}
%\subsection{Composition of the final catalogue}

The GLADE+ catalogue is available as a \texttt{txt} file on the GLADE website\footnote{GLADE website: \url{http://glade.elte.hu}}. Columns of the file contain the following data (where available) for each GLADE+ object:

\renewcommand{\labelenumi}{\arabic{enumi}}
\begin{enumerate}

    \item  GLADE+ catalogue number
	\item  Principal Galaxies Catalogue number
	\item  Name in the GWGC catalogue
	\item  Name in the HyperLEDA catalogue
	\item  Name in the 2MASS XSC catalogue
	\item  Name in the WISExSuperCOSMOS catalogue (wiseX)
	\item  Name in the SDSS-DR16Q catalogue
	\item  Object type flag: '$Q$' means that the source is from the SDSS-DR16Q catalogue, '$G$' means that it is from another catalogue and has not been identified as a quasar
	\item  Right ascension in degrees (J2000)
	\item  Declination in degrees (J2000)
			
	\item  Apparent $B$ magnitude
	\item  Absolute error of apparent $B$ magnitude
	\item  $B$ magnitude flag: '0' if the $B$ magnitude is measured, '1' if it is calculated from the $B_J$ magnitude
	\item  Absolute $B$ magnitude
	\item  Apparent $J$ magnitude
	\item  Absolute error of apparent $J$ magnitude
	\item  Apparent $H$ magnitude
	\item  Absolute error of apparent $H$ magnitude
	\item  Apparent $K_s$ magnitude
	\item  Absolute error of apparent $K_s$ magnitude
	\item  Apparent $W1$ magnitude
	\item  Absolute error of apparent $W1$ magnitude
	\item  Apparent $W2$ magnitude
	\item  Absolute error of apparent $W2$ magnitude
	\item  $W1$ flag: '0' if the $W1$ magnitude is measured, '1' if it is calculated from the $K_s$ magnitude
	\item  Apparent $B_J$ magnitude
	\item  Absolute error of apparent $B_J$ magnitude

	\item  Redshift in the heliocentric frame
	\item  Redshift converted to the Cosmic Microwave Background (CMB) frame
	\item  Redshift correction flag: '0' if the CMB frame redshift and luminosity distance values given in columns 29 and 33 are not corrected for the peculiar velocity, '1' if they are corrected values
	\item  Error of redshift from the peculiar velocity estimation
	\item  Measurement error of heliocentric redshift
	\item  Luminosity distance in Mpc units
	\item  Error of luminosity distance in Mpc units
	\item  Redshift and luminosity distance measurement flag: '0' if the galaxy has no measured redshift or distance value, '1' if it has a measured photometric redshift from which we have calculated its luminosity distance, '2' if it has a measured luminosity distance value from which we have calculated its redshift, '3' if it has a measured spectroscopic redshift from which we have calculated its luminosity distance
	%\item  Redshift error flag: '1' if the redshift error is a relative error of 36 percent, '2' if the redshift error is an absolute error of $1.5\times 10^{-4}$, '3' if the redshift error is an absolute error of $1.5\times 10^{-2}$, '4' if the redshift error is a relative error of 3.3 percent. Note that this does not include the potential redshift error from the peculiar velocity estimation, which is given in column 27
			
	\item  Stellar mass in $10^{10}\ M_{\odot}$ units
	\item  Absolute error of stellar mass in $10^{10}\ M_{\odot}$ units
	\item  Stellar mass flag: '0' if the stellar mass was calculated assuming no active star formation, '1' if the stellar mass was calculated assuming active star formation
	\item  Base-10 logarithm of estimated BNS merger rate in the galaxy in Gyr$^{-1}$ units
	\item  Absolute error of estimated BNS merger rate in the galaxy
\end{enumerate}

\section{CONCLUSIONS}
\label{sec:conclusion}

The GLADE+ galaxy catalogue is an extended version of the GLADE catalogue we have optimized for multimessenger searches with advanced GW detectors. It contains more than 23 million objects from which more than 22 million are galaxies and $\sim$750,000 are quasars. As the effects of peculiar motions are important for nearby galaxies for both the EM follow-up and cosmological analyses, we have estimated the peculiar velocities along with their standard deviations using the \borg ~forward modelling method for galaxies in GLADE+ with $B$-band magnitude data and having redshifts $z\leq 0.05$. GLADE+ is complete up to $d_L=47^{+4}_{-2}$~Mpc in terms of the cumulative $B$-band luminosity of galaxies, and contains all of the brightest galaxies giving half of the total $B$-band ($K_s$-band) luminosity up to $d_L\simeq 250$ Mpc ($d_L\simeq 390\ \mathrm{Mpc}$).

As according to theoretical models the stellar masses of galaxies strongly correlate with the merger rates of colliding binaries, we have calculated the stellar masses and the BNS merger rates (together with their errors) of each galaxy having WISE magnitudes. These parameters can help in ranking the galaxies in a given GW localization volume for EM follow-up observations, thereby possibly reducing the number of pointings and the total integration time needed to find the EM counterpart.

\section*{Acknowledgements}
This paper was reviewed by the LIGO Scientific Collaboration under LIGO Document P2100375.
The authors would like to thank Simone Mastrogiovanni, Surhud More and John Peacock for fruitful discussions throughout the project. The authors thank Bence Bécsy and Daniel Holz for useful comments on the manuscript. We are grateful for the Wide Field Astronomy Unit (WFAU) for providing the WISExSCOS and 2MPZ data used in creating GLADE+. We acknowledge the usage of the HyperLeda database (\url{http://leda.univ-lyon1.fr}). GD is supported through the \'UNKP-19-3 New National Excellence program of the Hungarian Ministry of Human Capacities and the iBOF-project BOF20/IBF/124. Research at Perimeter Institute is supported in part by the Government of Canada through the Department of Innovation, Science and Economic Development Canada and by the Province of Ontario through the Ministry of Colleges and Universities. The work of BDW is supported by the Labex ILP (reference ANR-10-LABX-63) part of the Idex SUPER, received financial state aid managed by the Agence Nationale de la Recherche, as part of the programme Investissements d'avenir under the reference ANR-11-IDEX-0004-02. The Center for Computational Astrophysics is supported by the Simons Foundation. This work was supported by the ANR BIG4 project, grant ANR-16-CE23-0002 of the French Agence Nationale de la Recherche. A part of the analysis was carried out at the Horizon cluster hosted by Institut d'Astrophysique de Paris. We thank Stephane Rouberol for smoothly running the Horizon cluster. This work was granted access to the HPC resources of CINES (Centre Informatique National de l'Enseignement Sup\'erieur) under the allocation A0020410153 made by GENCI. This work is done within the Aquila Consortium\footnote{\url{https://www.aquila-consortium.org/}}. JJ acknowledges support by the Swedish Research Council (VR) under the project 2020-05143 -- "Deciphering the Dynamics of Cosmic Structure". We acknowledge the use of following packages in this analysis: Astropy \citep{2013A&A...558A..33A, 2018AJ....156..123A}, IPython \citep{PER-GRA:2007}, Matplotlib \citep{Hunter:2007}, NumPy \citep{2011CSE....13b..22V}, and SciPy \citep{scipy}.\\

\section*{Data availability}
The data underlying this article are available at the GLADE website, \url{http://glade.elte.hu/}.

\bibliographystyle{mnras}
\bibliography{cites}

\appendix

\bsp
\label{lastpage}
\end{document}